\documentclass[sigconf]{acmart}
\usepackage{multirow}
\usepackage{caption}
\usepackage{subcaption}
\usepackage{amsmath}
\usepackage{hyperref}

\AtBeginDocument{%
  \providecommand\BibTeX{{%
    \normalfont B\kern-0.5em{\scshape i\kern-0.25em b}\kern-0.8em\TeX}}}

\copyrightyear{2022}
\acmYear{2022}
\setcopyright{rightsretained}
\acmConference[WWW '22 Companion]{Companion Proceedings of the Web Conference 2022}{April 25--29, 2022}{Virtual Event, Lyon, France}
\acmBooktitle{Companion Proceedings of the Web Conference 2022 (WWW '22 Companion), April 25--29, 2022, Virtual Event, Lyon, France}
\acmDOI{10.1145/3487553.3524674}
\acmISBN{978-1-4503-9130-6/22/04}

\begin{document}

\title[Towards Analyzing the Bias of News Recommender Systems]{Towards Analyzing the Bias of News Recommender Systems Using Sentiment and Stance Detection}

\author{Mehwish Alam}
\authornote{First two authors contributed equally to this research.}
\email{mehwish.alam@kit.edu}
\affiliation{%
  \institution{FIZ Karlsruhe, Karlsruhe Institute of Technology}
  \country{Germany}
}

\author{Andreea Iana}
\authornotemark[1]
\email{andreea@informatik.uni-mannheim.de}
\affiliation{%
  \institution{University of Mannheim}
  \country{Germany}
}

\author{Alexander Grote}
\email{alexander.grote@kit.edu}
\affiliation{%
  \institution{FZI Forschungszentrum Informatik, Karlsruhe Institute of Technology}
  \country{Germany}
}

\author{Katharina Ludwig}
\email{kaludwig@mail.uni-mannheim.de}
\affiliation{%
  \institution{University of Mannheim}
  \country{Germany}
}

\author{Philipp Müller}
\email{p.mueller@uni-mannheim.de}
\affiliation{%
  \institution{University of Mannheim}
  \country{Germany}
}

\author{Heiko Paulheim}
\email{heiko@informatik.uni-mannheim.de}
\affiliation{%
  \institution{University of Mannheim}
  \country{Germany}
}

\renewcommand{\shortauthors}{Alam and Iana, et al.}

\begin{abstract}
  News recommender systems are used by online news providers to alleviate information overload and to provide personalized content to users. However, algorithmic news curation has been hypothesized to create filter bubbles and to intensify users' selective exposure, potentially increasing their vulnerability to polarized opinions and fake news. 
  In this paper, we show how information on news items' stance and sentiment can be utilized to analyze and quantify the extent to which recommender systems suffer from biases. To that end, we have annotated a German news corpus on the topic of migration using stance detection and sentiment analysis.
  In an experimental evaluation with four different recommender systems, our results show a slight tendency of all four models for recommending articles with negative sentiments and stances against the topic of refugees and migration. Moreover, we observed a positive correlation between the sentiment and stance bias of the text-based recommenders and the preexisting user bias, which indicates that these systems amplify users' opinions and decrease the diversity of recommended news. The knowledge-aware model appears to be the least prone to such biases, at the cost of predictive accuracy.
\end{abstract}



\maketitle

\section{Introduction}

Every day, large volumes of news articles are being published online, leading to an information load that exceeds the consumptive capacities of Internet users. To alleviate this information overload and to provide customized content to users based on their past interests, recommender systems are widely employed by news providers. However, this algorithmic news curation influences users' exposure to diverse content, by selectively filtering out articles that seem irrelevant (in order to maximize user engagement) \cite{pariser2011filter} or inconsistent with the readers' preexisting beliefs and attitudes, as users are more likely to accept information that reinforces their opinions \cite{freedman1965selective}. The power of recommender systems to shape users' perception of the world has over time lead to concerns that individuals are being isolated from diverse perspectives through algorithmically created "filter bubbles" \cite{pariser2011filter} and self-selected "echo chambers" \cite{sunstein2001echo,jamieson2008echo}, in which they only interact with individuals with similar ideological viewpoints. In the context of (political) news, an over-exposure to less diverse viewpoints may determine an attitude-reinforcing spiral \cite{stroud2017selective,donsbach2013dissonant} which, in the long run, can lead to opinion polarization, or even radicalization of individuals with extreme political or ideological viewpoints \cite{bakshy2015exposure,boutyline2017social,liu2021interaction}.

Diversity represents a paramount characteristic not only for news quality \cite{mcquail1992media}, but also for ensuring a balanced and broad variety of information to create a well-informed public in any democratic society \cite{helberger2019democratic}. Nonetheless, news diversity is a multi-faceted concept \cite{haim2018burst} which can refer, simultaneously, to pluralism of sources (\textit{source diversity}) \cite{voakes1996diversity}, of discussed topics (\textit{content diversity}) \cite{haim2018burst}, and of stances taken on a given topic (\textit{viewpoint diversity}) \cite{baden2017conceptualizing}. Some previous works have focused on constructing datasets for detecting and analyzing the more general notion of media bias in news articles \cite{spinde2020integrated,lim2020annotating,farber2020multidimensional}, and on establishing a connection between the research on media bias in the social science and computer science fields \cite{hamborg2019automated}. 

From a recommender-system point of view, sentiment analysis and stance detection have been used to measure and control news diversity. In sentiment analysis, systems determine whether a piece of text is positive, negative, or neutral \cite{liu2010sentiment,ravi2015survey}. However, in stance detection, systems are to determine an article’s or author's viewpoints towards a given target issue, which may not even be explicitly mentioned in the text \cite{KucukC20,reuver2021no}. Stance and sentiment annotations can be used to identify whether news recommender systems intensify selective exposure and increase users' vulnerability to polarized opinions and fake news. At the same time, stance and sentiment labels can constitute cues for controlling the level of polarization in personalized recommender systems, and for generating suggestions with more diverse sentiment orientations and viewpoints.

In this paper, we use sentiment and stance annotations to analyze whether different kinds of recommender systems suffer from an underlying bias towards a certain sentiment or viewpoint, and whether this decreases diversity of recommendations and intensifies users' selective exposure~\cite{abs-2104-05994}. To this end, we use transfer learning to annotate GeNeG, a dataset of German news articles, with sentiment scores and stance labels. These annotations are then used to examine the results of four recommender systems with regards to the sentiment and stances identified in the suggested articles, as well as those from the user's reading history. 

The rest of the paper is structured as follows. Firstly, we discuss related work on stance detection in various languages (Section~\ref{sec:related_work}). Secondly, we introduce GeNeG and describe its curation and annotation process (Section~\ref{sec:geneg}). Thirdly, in Section \ref{sec:analysis} we evaluate a set of news recommender systems, and examine their sentiment and stance bias. We conclude with a discussion of the findings and an outlook of the next steps planned for this study in Section~\ref{sec:discussion}.

\section{Related Work}
\label{sec:related_work}

In the following, various studies are discussed which take into account the creation of datasets for stance detection in different languages since the multilinguality aspect of stance detection is fairly understudied. We then more particularly focus on the multilinguality aspect in stance detection. For an exhaustive survey on techniques for stance detection please refer to ~\cite{KucukC20}. In the current study, we do not discuss knowledge-aware news recommender systems, however, for more details on knowledge-aware news recommender systems please refer to \cite{IanaAP21}.

\subsection{Existing Datasets for Stance Detection}

STANDER~\cite{ConfortiBPGTC20} is a dataset for stance detection and fine-grained evidence retrieval task for future research in stance detection, as well as multitask learning. It contains 3,291 annotated articles by experts (professional journalists). This corpus is also aligned with the Twitter dataset WT-WT \cite{ConfortiBPGTC20} corpus. The topic of the news articles in STANDER is ``mergers of US companies in the healthcare industry". The stance annotations are `support', `refute', `comment', and `unrelated'. Moreover, the corpus also includes `evidences' as an annotation which indicates the part of text used to identify the stance of the article. 2 to 4 annotators were employed, and majority voting was used to decide on the final annotation.

In~\cite{MohammadKSZC16}, the authors present a dataset of tweets annotated with the tweeter's stance (in favor or against) regarding an already chosen target. The dataset contains six targets of interest commonly debated in the United States, i.e, ‘Atheism’, ‘Climate Change is a Real Concern’, ‘Feminist Movement’, ‘Hillary Clinton’, ‘Legalization of Abortion’, and ‘Donald Trump’. For each target, 1000 tweets were randomly sampled from the initially collected 1.7 million tweets. The tweets were annotated by 8 annotators. In total, the corpus contains 4,870 annotated tweets. This dataset has been used for the stance detection task in SemEval 2016\footnote{\url{https://alt.qcri.org/semeval2016/task6/}}. This stance dataset, which was subsequently also annotated for sentiment, can be used to better understand the relationship between stance, sentiment, entity relationships, and textual inference.

In the Fake News detection Challenge, FNC-1~\cite{HanselowskiSSCC18}, the organizers focus on performing stance detection on document level. The documents are to be classified into four classes: ‘agree’, ‘discuss’, ‘disagree’, ‘unrelated’. The data contains 2,587 headlines and documents. Out of these documents, 7.4\%  are labeled ‘agree’, 2.0\% ‘disagree’, 17.7\% ‘discuss’, and 72.8\% ‘unrelated’, respectively. All the above datasets for stance detection are for English language, and thus, not usable for performing stance detection on GeNeG.

\subsection{Multilingual Stance Detection}

In \cite{mascarell-etal-2021-stance} the authors create Cheese, a new dataset containing 3,693 pairs of debate questions and the associated Swiss news articles in German, annotated with stances and emotions. The news articles were published between 2004 and 2020, and cover 24 topics, including science, environment, politics, religion, society, etc. The stance annotation include ‘in favor’, ‘against’, ‘discussion’, and ‘unrelated’, while the emotion annotations are ‘joy’, ‘trust’, ‘fear’, ‘anticipation’, ‘sadness’, ‘disgust’, ‘anger’, ‘surprise’, and 'no emotion'. 
Along with the dataset, the authors provide a supervised classification task targeting the stance of the news article with respect to the question. The classification algorithm uses a contextual language model in German (German BERT).

SardiStance~\cite{CignarellaLBPR20} is a stance detection task for tweets in Italian. The shared task was created for two different settings: (i) textual stance detection, exploiting only the information provided by the tweet, and (ii) contextual stance detection, with the addition of information on the tweet itself, such as the number of retweets, the number of favors or the date of posting, contextual information about the author, such as follower count, location, user’s biography, and additional knowledge extracted from the user’s network of friends, followers, retweets, quotes and replies. The dataset constructed for the task contains 700K tweets written in Italian about ``Movimento delle Sardine``. The tweets were collected from 11-2019 to 01-2020, and annotated with the labels ‘against’, ‘in favor’, ‘neutral’, and ‘irrelevant’. The dataset also contains annotations indicating whether the tweet is ‘ironic’, ‘non-ironic’, or ‘undefined’.

In~\cite{XuZWGDX16}, the authors present the task of stance detection on Chinese Microblogs. This is further divided into two sub-tasks, where the first is a supervised classification task which detects the stance towards five targets of interest with given labeled data. The second is an optional unsupervised task which requires only unlabeled data.

Previously discussed datasets focus on only one language. As a step further, in x-stance~\cite{corr/abs-2003-08385} the authors facilitate multilingual, multi-target stance detection, which also helps in performing cross-lingual stance detection. The authors use Multilingual BERT~\cite{DevlinCLT19} for performing stance classification. The dataset contains question and news article pairs in English, French, German, and Italian.  

\section{Enriching German News Knowledge Graph with Sentiments and Stance}
\label{sec:geneg}

This section discusses necessary details about GeNeG, along with the methods adopted for performing sentiment analysis as well as stance detection.

\subsection{GeNeG}
\label{subec:geneg_description}
GeNeG~\cite{andreea_iana_2022_5913171} is a news knowledge graph constructed from a dataset of 4,557 German news articles on the topic of refugees and migration, collected from 39 media outlets that cover a wide political spectrum, including far-right and far-left online publications. The corpus consists of news published between 01-01-2019 and 20-10-2020, selected based on keyword stems representative of the topic (e.g., ‘flüchtl’, ‘migrant’, ‘asyl’), and collected from the news outlets via a Web crawler\footnote{\url{https://github.com/andreeaiana/german-news}}. The dataset contains rich textual and metadata information, such as content, provenance, publishing dates, authors, or tagged keywords. Furthermore, named entities have been extracted from the articles' content (e.g. persons, locations) and metadata (e.g. publisher, author) and disambiguated using Wikidata \cite{vrandevcic2014wikidata}. 

The news knowledge graph built from this dataset represents a heterogeneous network, where the news content and real-world entities are represented as nodes, while different relations between these items constitute the graph's edges. The nodes are divided into \textit{literals}, denoting textual content, dates or polarization scores, and \textit{entities}, indicating identified named entities. In turn, entity nodes can be linked to Wikidata, or non-disambiguated (if they are not found in Wikidata). GeNeG is provided in three flavors: a \textit{base graph} -- containing textual information, metadata, and entities extracted from the articles, an \textit{entities graph} -- derived from the base version by removing all literal nodes, it contains only entities extracted from the articles and it is enriched with their three-hop Wikidata neighbors, and a \textit{complete graph} -- combining the previous two graphs, and incorporating both literals and entities. The \textit{base GeNeG} contains 54,327 nodes and 186,584 edges denoting 16 properties. The \textit{entities GeNeG} consists of 844,935 nodes, and 6,615,972 edges representing 1,263 properties. Lastly, the \textit{complete GeNeG} comprises of 868,159 nodes and 6,656,779 edges denoting 1,271 properties.

\subsection{Sentiment Annotations}
\label{subsec:sentiment_annotation}

To annotate each news article with a sentiment score, we used a pre-trained BERT-based \cite{DevlinCLT19} sentiment classification model for German language texts \cite{guhr-etal-2020-training}. The model uses a softmax function to calculate a probability estimate, which classifies each document as either positive, neutral or negative. To transform these probabilities into a sentiment score in the range of -1 to 1, we take the negative sentiment probability and subtract it from the positive sentiment probability. More formally, this is expressed by \(s=p_p-p_n\), where $s$ equals the sentiment score, while \(p_p\) and \(p_n\) represent the positive and negative sentiment probability, respectively. We ignore the neutrality score, since it is implicitly encoded as \(1-p_p-p_n\). On the GeNeG corpus we observed a slight skewness towards articles with negative sentiment, as indicated by the mean and median sentiment scores of $s_{mean}=-0.154$, and $s_{median}=-0.019$, respectively.

\subsection{Transfer Learning for Stance Detection on German News Articles}

In order to perform stance detection on the news articles collected on the topic of refugees and migration, transfer learning was performed. In order to do so, the first step is to select a proper training dataset for articles in German, and the second step is to classify the articles into two classes, namely 'in favor ' or 'against'.

\subsubsection*{{\bf Training Data Selection}}

There are two datasets which are introduced for German language, i.e., Cheese~\cite{mascarell-etal-2021-stance} which specifically focuses on German language, and the German part of x-stance~\cite{corr/abs-2003-08385}. The Cheese dataset was excluded from consideration for training since the number of instances in each of the classes, i.e., ‘in favor’, ‘against’, ‘unrelated’, ‘discussion’, were highly unbalanced (i.e., 702, 286, 1428, and 774), leading to unbalanced classes predicted for our corpus. This issue has been discussed in~\cite{corr/abs-2007-15121} in more detail, where the authors propose a cascade of binary classifiers to consider the hierarchy of classes in stances. More specifically, the authors first classify the text into relevant or irrelevant, and then label the relevant texts as either 'neutral' or 'stance'. The final binary classification is performed using 'in favor' or 'against' class labels for the texts which were previously classified as 'stance'. 

In our case, we use the x-stance~\cite{corr/abs-2003-08385} dataset for training the classifier, which contains a balanced number of articles for both classes, i.e., ‘in favor’ and ‘against’. The training data was extracted from x-stance with the language label 'de', i.e., only the German subset. The statistics of the corpus extracted from x-stance are given in Table~\ref{tab:x-stance-stats}.

\begin{table}[!htb]
    \centering
    \caption{Statistics of the the German part of the x-stance dataset.}

    \begin{tabular}{|l|r|r|r|}
    \hline
     Dataset & Against & Favor & Total \\ \hline
     Training & 17130 & 16720 & 33850\\ \hline
     Validation & 1451 & 1420 & 2871\\ \hline
     Test & 5882 & 6009 & 11891\\ \hline
      
    \end{tabular}
    \label{tab:x-stance-stats}
\end{table}

\subsubsection*{{\bf Stance Detection in GeNeG}}

Stance detection takes into account a question and article pair, and classifies the stance of the article as being in favor or against the question at hand. The questions used to build the question and news article pairs for classification are given in Table~\ref{tab:questions}. GermanBert\footnote{\url{https://huggingface.co/bert-base-german-cased}} \cite{chan-etal-2020-germans} was fine-tuned on the training dataset provided in x-stance.  The hyper-parameters for the best performing model on the test set are: $learning~rate = 3e-5$ and  $number~of~epochs = 4$. Afterwards, transfer learning was performed for classifying the news articles in GeNeG (as test dataset) into one of the two classes, i.e., ‘in favor’ or ‘against’, according to one of the predefined questions. In future studies we would like to perform human evaluation on the annotated GeNeG articles. Table~\ref{tab:stance-detection-results} show the results of the stance detection for each of the questions, as well as the average score, computed as $\frac{favor-against}{favor+against}$.

\begin{table*}[!htb]
    \centering
    \caption{Questions for the question-news article pairs.}

    \begin{tabular}{|l|l|}
    \hline
    German Question & English Translation (for understandability) \\ \hline
    (Q1) Befürworten Sie, dass Flüchtlinge nach Deutschland kommen? &  Are you in favor of refugees coming to Germany? \\ \hline
    (Q2) Befürworten Sie, dass Flüchtlinge in Deutschland leben? & Are you in favor of refugees living in Germany? \\ \hline  
    (Q3) Befürworten Sie, dass Flüchtlinge in Deutschland arbeiten? &  Are you in favor of refugees working in Germany?  \\ \hline 
    (Q4) Sollte Deutschland Flüchtlinge aufnehmen? & Should Germany take in refugees?  \\ \hline
    (Q5) Sollte Deutschland Flüchtlingen helfen? & Should Germany help refugees?  \\ \hline
      
    \end{tabular}
    \label{tab:questions}
\end{table*}

\begin{table}[!htb]
    \centering
    \caption{Articles ‘in favor’ of or ‘against’ the pairing question.}

    \begin{tabular}{|l|r|r|r|}
    \hline
    Question & Articles in Favor & Articles Against & Avg. Score\\ \hline
    (Q1)  &2165 & 2392  & -0.050\\ \hline
    (Q2)  & 2193  & 2364 & -0.038\\ \hline  
    (Q3)  & 2210 & 2347 & -0.030\\ \hline 
    (Q4) & 2120   & 2437 & -0.070\\ \hline
    (Q5) & 2192  &  2365 & -0.038\\ \hline
      
    \end{tabular}
    \label{tab:stance-detection-results}
\end{table}

Finally, GeNeG was additionally populated with the output of the stance detection with the help of two newly created properties, namely {\tt geneg:in\_favor} and {\tt geneg:against}. The resulting triples have the form $\langle article, stance, question \rangle$, where $stance \in \{ {\tt geneg:in\_favor},  {\tt geneg:against}\}$, and $question \in \{Q_n, n=\overline{1,5}\}$. The dataset is available through Zenodo\footnote{\url{https://doi.org/10.5281/zenodo.6039372
}} with restricted access.

\section{Sentiment and Stance Bias of News Recommenders}
\label{sec:analysis}

The stance and sentiment of news articles can be used to determine whether news recommenders are biased towards a certain sentiment or stance of the articles, and if this, in turn, correlates with and reinforces users' existing viewpoints on a given topic. In this section, we firstly introduce the four news recommender systems analyzed and describe the user data. Afterwards, we evaluate the performance of the models and investigate their sentiment and stance bias.

\subsection{Recommendation models}

Collaborative filtering is the most adopted recommendation method in fields such as music or movie recommendation. Nonetheless, content-based approaches are the most widely used in the field of news recommendation mostly due to the fact that users generally do not have long-term profiles on news websites, with reading history being limited to a single session, and feedback being collected almost exclusively implicitly from click logs \cite{karimi2018news,IanaAP21}. 
We compared four different content-based recommender systems in our analysis. The first three are \emph{textual}, i.e., they are solely based on the articles' texts, whereas the last one is \emph{knowledge-based}, using the graph representation and the information from Wikidata instead of the articles' texts. All text-based models use cosine similarity to determine the similarity between a candidate news article and the user's history of read news. 

\begin{itemize}
    \item The \textbf{Term Frequency-Inverse Document Frequency (TF-IDF)} recommender uses a TF-IDF \cite{schutze2008introduction} vector representation for each article.
    \item The \textbf{Word2vec} recommender encodes articles using word embeddings learned from a large text corpus \cite{mikolov2013efficient}. More specifically, we use a pre-trained German word2vec model trained on the Common Crawl and Wikipedia dataset \cite{grave2018learning} to learn latent representations of the words in an article. The article's vector representation is computed as the average of its words' embeddings.
    \item The \textbf{Transfomer} architecture uses an attention mechanism to incorporate context in text embeddings \cite{vaswani2017attention}. We use a pre-trained cross-lingual model for sentence embeddings in English and German \cite{reimers2019sentence,cross-en-de-robert} to encode the sentences in the news articles. The final representation of the article is obtained by averaging the embedding vectors of all of its sentences. 
    \item \textbf{RippleNet} \cite{wang2018ripplenet} is a knowledge-aware recommender which propagates a user's potential preferences along the edges of the knowledge graph. The model generates \textit{ripple sets} (i.e. sets of multi-hop entity neighbors encoding potential user interests) based on the entities extracted from the user's read articles. These ripple sets are used to explore higher-order preferences, where the strength of the user's preference diminishes proportional to the distance from the original seed in the knowledge graph. The final click probability is predicted using the preference distribution of the user for a candidate news, obtained by superposing multiple ripple sets. For RippleNet, we use the \textit{entities GeNeG} as the knowledge graph for the recommender. In particular, the sentiment and stance information was \emph{not} included in the graph used to compute the recommendations.
\end{itemize}

\subsection{User Data}
We collected the user data through an online study aimed at measuring the political polarization effect of recommender systems on users. The study was based on the news dataset described in Section~\ref{subec:geneg_description}. However, 732 articles longer than 1,500 words were removed from the experiment in order to limit the reading time required for the participants, and to ensure a strong response to the stimulus. Each of the participants in the study was randomly assigned to one of four recommenders, namely TF-IDF, Word2vec, Transformer, or a random recommendation baseline. Each participant was asked to choose an article from a preview of six articles, and then to read it. The user's choices were included in his or her reading history. This process was repeated four times, resulting in four interactions per participant. 

We split the user data into training and test sets using a 80:20 ratio. Since 85\% of the user ratings were generated using recommendations of the text-based recommenders, these baselines are prone to overfit. Therefore, we generate a test subset containing only user ratings for randomly suggested articles from the complete test set. The user data statistics can be found in Table \ref{tab:user_data_stats}. For more details on the online study and data collection, please refer to Appendix \ref{sec:user_study}.

\begin{table}[!htb]
    \centering
    \caption{User data statistics.}
    \begin{tabular}{|l|r|r|} 
    \hline
    Dataset & Items & Users \\ \hline
    Total & 3,825 & 1,417 \\ \hline
    Training & 3,365 & 1,414 \\ \hline
    Complete test & 1,633 & 1,174 \\ \hline
    Random test & 316 & 177 \\ \hline
    \end{tabular} %
    \label{tab:user_data_stats}
\end{table}

\subsection{Evaluation of Recommender Systems}

In the following, we describe the experimental setup and discuss the evaluation results of the proposed recommenders.

\subsubsection{Experimental Setup}

We evaluated the four recommender models on click-though-rate (CTR) prediction. In this scenario, each recommender is applied on every user-article pair from the test set to predict the user's likelihood of clicking the candidate article. Furthermore, we applied a min-max scaling to the similarity measures outputted by the text-based recommenders as an approximation of probability scores. We use Accuracy (ACC), Area Under the Curve (AUC), and the F1 score to evaluate the performance of the models. 

The key parameters settings for the analyzed recommender systems \footnote{The code and data are available at https://github.com/andreeaiana/geneg\_benchmarking.} are as follows. For TF-IDF, we use an n-gram range of 1 and 2, and l2-norm regularization. The Word2vec and the Transformer recommenders use embedding vectors of dimension 300, and respectively, 768. In RippleNet, we empirically set the number of hops to $H=1$, the size of the user's ripple set to 16, the dimension of the item and knowledge graph embeddings to $d=48$, and the training weight of the knowledge graph embedding to $\lambda_2=0.03$. We use the default values for news recommendation from \cite{wang2018ripplenet} for the other parameters.

\subsubsection{Results}

The experimental results of the benchmarked recommenders are summarized in Table \ref{tab:ctr_results}. The text-based recommenders achieve the best scores in terms of AUC and F1 on the complete test set, whereas RippleNet outperforms the Word2vec and Transformer recommenders in terms of accuracy. However, the text-based recommendation models overfit on the complete test data, due to the user ratings provenance. More specifically, the ratings collected from users assigned to a TF-IDF recommender account for more than 55\% of the cases (for more in-depth statistics regarding the ratings' provenance, please refer to Appendix \ref{sec:user_study}). 

\begin{table}[!htb]
    \caption{CTR prediction results.}
    \begin{tabular}{|l|r|r|r|r|r|r|}
    \hline
    \multirow{2}{*}{Model} & \multicolumn{3}{c|}{Complete} & \multicolumn{3}{c|}{Random} \\
    
    \multicolumn{1}{|c|}{} & \multicolumn{1}{c}{ACC} & \multicolumn{1}{c}{AUC} & \multicolumn{1}{c|}{F1} 
    & \multicolumn{1}{c}{ACC} & \multicolumn{1}{c}{AUC} & \multicolumn{1}{c|}{F1} \\\hline
    
    TF-IDF  & \textbf{0.732} & \textbf{0.873} & 0.647 
            & 0.487 & 0.499 & 0 \\ \hline
            
    Word2vec    & 0.514 & 0.794 & \textbf{0.674}
                & 0.499 & 0.474 & 0.663 \\ \hline
            
    Transformer & 0.505 & 0.779 & 0.671 
                & 0.499 & 0.515 & \textbf{0.665} \\ \hline
    
    RippleNet   & 0.553 & 0.574 & 0.523 
                & \textbf{0.559} & \textbf{0.578} & 0.531 \\ \hline
    \end{tabular} %
    \label{tab:ctr_results}
\end{table}

Consequently, we investigated the performance of the recommender systems on the random test set, in order to reduce the bias of the text-based models. In this evaluation setting, RippleNet outperforms all other recommenders in terms of accuracy and AUC. Moreover, it obtains a higher F1 score than the TF-IDF model,which is unable to make any correct predictions for the articles in this test set, resulting in a F1 score of 0. 

Overall, we conclude that, by using only entities in GeNeG for computing recommendations, RippleNet is able to achieve a decent performance, while being more robust to changes in the structure of the data compared to the purely text-based recommenders. 

\subsection{Bias Analysis}

In addition to the recommenders' predictive accuracy, we analyzed their recommendations to identify whether the recommenders are prone to stance or sentiment bias. In this study, each of the text-based models generated an output containing the top $k$ most similar articles to the ones in the user history, whereas RippleNet recommended the $k$ articles with the highest probability of being clicked by the user. In our subsequent bias analysis, we set the number of recommended articles for all recommenders to $k=5$, as we observed that the majority of investigated outlets suggest, on average, five other articles related to the one currently read by the user.

In order to quantify the stance of the news articles and to calculate an overall bias score for the users and recommenders, we employ the following transformation function for an article's stance label:

\[
 stance\_score = 
  \begin{cases} 
  +1 & \text{if } stance\_label = \text{Favor}\\
  -1 & \text{if } stance\_label = \text{Against}\\
  \end{cases}
\]

For each user, we compute the average sentiment bias score as the mean of the sentiment scores of the articles included in his or her reading history. Similarly, the recommender sentiment bias score per user was obtained as the average of all sentiment scores of the recommended articles. Lastly, the mean of all recommender sentiment bias scores over all users constitutes the recommender's average sentiment bias score. The average user and recommender stance bias scores are calculated analogously. The average bias scores for both sentiment and stances fall in the interval $[-1, 1]$, where -1 represents a user's or recommender's tendency for articles with negative sentiments or stances against the given topic, whereas +1 denotes the opposite situation.

\subsubsection{Recommender Bias}

\begin{table*}[!htb]
    \centering
    \caption{Average user and recommender sentiment scores. The statistical significance with a Student's $t$-test is denoted with * ($p-value < 0.01$) and ** ($p-value < 0.05$). A star in the case of user sentiments denotes statistical significance between the average user and corpus scores. In the case of recommender sentiment, (*/*) denotes statistical significance, firstly with the average user score, and secondly with the average corpus score. A ‘-’ sign shows that no such significance was found.}
    \begin{tabular}{|l|r|r|r|r|r|}
    \hline
    \multirow{2}{*}{Test set} &  \multirow{2}{*}{\begin{tabular}[c]{@{}c@{}}Avg. user\\ sentiment score\end{tabular}} & \multicolumn{4}{c|}{Avg. recommender sentiment score}\\ 
     & & TF-IDF & Word2vec & Transformer & RippleNet \\ \hline
    
    Complete & -0.171*  & -0.162 (-/-) & -0.169 (-/**) & -0.157 (-/-) & -0.148 (*/-)  \\ \hline
    Random &  -0.169 & -0.141 (-/-) & -0.170 (-/-) & -0.160 (-/-) & -0.150 (**/-) \\   \hline

    \end{tabular}
    \label{tab:avg-sentiment-bias}
\end{table*}

\begin{table*}[!htb]
    \centering
    \caption{Average user and recommender stance scores. The statistical significance with a Student's $t$-test is denoted with * ($p-value < 0.01$) and ** ($p-value < 0.05$). A star in the case of user stance denotes statistical significance between the average user and corpus scores. In the case of recommender stance, (*/*) denotes statistical significance, firstly with the average user score, and secondly with the average corpus score. A ‘-’ sign shows that no such significance was found.}
    \begin{tabular}{|l|r|r|r|r|r|}
    \hline
    \multirow{2}{*}{Question} & \multirow{2}{*}{{\begin{tabular}[c]{@{}c@{}}Avg. user\\ stance score\end{tabular}}} & \multicolumn{4}{c|}{Avg. recommender stance score (complete / random test set)}\\ 
     & & \multicolumn{1}{c|}{TF-IDF} & \multicolumn{1}{c|}{Word2vec} & \multicolumn{1}{c|}{Transformer} & \multicolumn{1}{c|}{RippleNet} \\ \hline
    
    (Q1) & -0.109 / -0.093  & -0.140 (-/-) / -0.227 (**/--) & -0.165 (**/-) / -0.219 (**/-) & -0.136 (-/-) / -0.172 (-/-) & -0.082 (-/-) / -0.054 (-/-) \\ \hline
    (Q2) &  -0.102** / -0.093 & -0.132 (-/-) / -0.220 (**/--) & -0.158 (**/-) / -0.207 (**/-) & -0.131 (-/-) / -0.169 (-/-) &  -0.074 (-/-) / -0.038 (-/-)\\   \hline
    (Q3) & -0.092** / -0.081  & -0.127 (-/**) / -0.215 (**/--) & -0.149 (**/**) / -0.205 (**/-) & -0.116 (-/**) / -0.164 (-/-) &  -0.062 (-/**) / -0.024 (-/-)\\  \hline
    (Q4) &  -0.117 / -0.106 & -0.167 (**/-) / -0.255 (**/--) & -0.178 (**/-) / -0.268 (*/-) & -0.157 (-/-) / -0.179 (-/-) &  -0.095 (-/-) / -0.084 (-/-)\\ \hline
    (Q5) &  -0.079 / -0.081 & -0.130 (**/-) / -0.237 (*/--) & -0.135 (**/-) / -0.199 (**/-) & -0.124 (-/-) / -0.143 (-/-) &  -0.060 (-/-) / -0.055 (-/-)\\ \hline

    \end{tabular}
    \label{tab:avg-stance-bias}
\end{table*}

Firstly, we used the average recommender bias scores to answer the following question: \textbf{do the recommender systems have a tendency to recommend articles with a certain sentiment or stance?} 

Table \ref{tab:avg-sentiment-bias} shows the results in terms of \textit{sentiment bias}. As it can be seen here, all recommenders are more likely to suggest articles with negative sentiments. However, this skewness towards negative sentiments should be interpreted by taking into account that the news articles in GeNeG have, on average, a negative sentiment score, as shown in Section \ref{subsec:sentiment_annotation}. Nonetheless, we observe that textual recommenders, in 5 out of 6 cases, are more prone towards news with negative sentiments, as indicated by the average recommender sentiment bias, which is larger than the average sentiment score of the articles in the dataset, namely $s_{mean}=-0.154$. In contrast, the knowledge-aware model recommends, in both cases, articles with slightly less negative sentiment than the average sentiment score of the news in the corpus. However, these observations are statistically significant only for the Word2vec recommender on the complete test set, for $p-value < 0.05$ , as indicated in Table \ref{tab:avg-sentiment-bias}.

Similarly, Table \ref{tab:avg-stance-bias} illustrates the average recommender \textit{stance bias} on both test sets. All recommenders show a bias towards news articles that take a stance against the topic of refugees and migration, for all questions used to represent this topic. As shown in Table~\ref{tab:stance-detection-results}, in GeNeG there are more articles against the topic, than in favor of it, for all of the five questions used in the process of stance detection. While this can partly explain the recommenders' preferences for the more negative articles, the average scores for the articles selected by the recommenders are still more negative than the average scores in the entire corpus depicted above. Again, as for the sentiment bias, the knowledge-based recommender has the weakest negative bias, showing only little deviation from the overall average scores in the dataset. Nonetheless, the difference between the average sentiment scores of the recommenders and the corpus is statistically significant only for question (Q3).

\subsubsection{Correlation between the Recommender and User Biases}

We then pose the question: \textbf{how does the recommenders' sentiment and stance bias correlate with the existing user sentiment and stance bias?} This analysis aims to identify whether the recommenders tend to generate suggestions that agree with or reinforce the user's existing beliefs. Tables \ref{tab:avg-sentiment-bias} and \ref{tab:avg-stance-bias} show the average sentiment and stance user biases. 

We observed that readers have a slight preference for news articles with negative sentiments (Table \ref{tab:avg-sentiment-bias}), which is not fully explained by the overall skewness towards negative sentiment in the dataset. Nevertheless, in seven out of eight cases, the recommender's average sentiment bias is marginally less negative than the user bias, indicating that the recommenders include articles with less negative sentiment in their results list. Here again, one should take into consideration that the articles in our dataset have, on average, a negative sentiment. Even in this context, we notice that users are more prone to reading articles with negative sentiments, with the average user bias on the complete test set being -0.171, and respectively, -0.169 on the random test set, slightly larger than the average sentiment score of the news articles in GeNeG, $s_{mean}=-0.154$. According to the Student's $t$-test, this is a statistically significant difference. However, there does not seem to be a large difference between the recommender and the users' bias scores, with the former following the same pattern. This could indicate that the recommenders indeed learn from the user's history and preferences and generate suggestions that amplify the readers' existing preferences for a given sentiment. However, we observed a statistically significant difference between the user and the recommender average sentiment scores in the case of RippleNet on the complete test set, which shows that the knowledge-based recommender might be less prone to amplifying existing sentiment preferences. 

A similar pattern can be observed in the case of stance bias. On the one hand, the text-based recommenders appear to exacerbate the user's preference towards news against the topic of refugees and migration, as indicated by the larger negative average stance bias scores of the recommenders, on both test sets. Additionally, this behavior of the recommenders appears to be more pronounced on the random test set. The difference between the average stance scores of Word2vec recommendations and the average user stance scores are significant on both test sets, while in the case of TF-IDF this observation holds true mostly on the random test set. On the other hand, although the knowledge-aware recommender also seems biased towards viewpoints against the given topic, it also appears to suggest, on average, more articles which also have favorable stances towards the topic. This is shown by the average stance bias score of RippleNet, which is lower than the average user stance bias score. We observed no statistically significant differences between the user and recommender stance bias in the case of the Transformer and RippleNet recommenders.

Through a closer look at the correlation between the recommender and the user sentiment and stance biases, we identified five possible cases of bias correlation, as follows:

\begin{itemize}
    \item (C1) Bias in the same direction: both the user and the recommender have a tendency towards negative (positive) sentiments and/or against (in favor) stances.
    \item (C2) Bias in opposite directions: the user has a preference for negative (positive) sentiments and/or against (in favor) stances, while the recommender tends to suggest articles with an opposite sentiment and/or stance. 
    \item (C3) Skewed towards the user: the user has a preference for articles with negative (positive) sentiments and/or against stances, while the recommender is balanced with regards to the sentiment and/or stance of the recommended articles. 
    \item (C4) Skewed towards the recommender: the user is balanced with regards to the sentiment and/or stance of the read articles, while the recommender tends to suggest articles with negative (positive) sentiment and/or against (in favor) stance.
    \item (C5) No bias: neither the user, nor the recommender show any sentiment and/or stance bias. 
\end{itemize}

It should be noted that in the case of sentiment bias, only the first two cases, (C1) and (C2), are possible, given the nature of the sentiment scores, and the fact that there are no perfectly neutral articles in the corpus. 

Table \ref{tab:counts-sentiment-bias} shows the number of users that fall in each of the first two cases with regards to \textit{sentiment bias}. The results show that for both test sets, almost always the recommender will suggest articles with the same sentiments as the ones found in the user's history. 

\begin{table}[!htb]
    \centering
    \caption{Recommender-user sentiment correlation counts.}
    \begin{tabular}{|l|r|r|r|r|}
    \hline
    \multirow{2}{*}{Bias case} & \multicolumn{4}{c|}{Counts (complete / random test set)}\\ 
      & \multicolumn{1}{c|}{TF-IDF} & \multicolumn{1}{c|}{Word2vec} & \multicolumn{1}{c|}{Transformer} & \multicolumn{1}{c|}{RippleNet} \\ \hline
    
    (C1) & 1166 / 177 & 1166 / 177 & 1165 / 177  & 1160 / 177 \\ \hline
    (C2) &   8 / 0 & 8 / 0 & 9 / 0 &  14 / 0 \\   \hline

    \end{tabular}
    \label{tab:counts-sentiment-bias}
\end{table}

Furthermore, Table \ref{tab:counts-stance-bias} shows the number of users for each of the five bias cases and questions in terms of \textit{stance bias}. We observe that all recommenders generate recommendations that most often fall in one of the first two cases, with the same-direction bias being more likely. The third most probable case is the one in which the overall stance bias score of the user is close to 0, denoting a balanced news consumption with articles both in favor and against the topic, whereas the recommender is prone to suggesting articles with one particular viewpoint. The least likely cases are the ones in which the user is biased towards a certain stance, while the recommender is neutral, followed by the scenario in which both the user and the recommender are neutral with regards to the stance of the news. Overall, the prevalence of the the first bias case is indicative of the fact that recommenders not only learn from user's reading history, but also tend to amplify existing opinions by recommending news with the same stances as the ones previously read by the user.

\begin{table}[!htb]
    \centering
    \caption{Recommender-user stance correlation counts.}
    \resizebox{\columnwidth}{!}{%
    \begin{tabular}{|l|l|r|r|r|r|}
    \hline
    \multirow{2}{*}{Question} & \multirow{2}{*}{Bias case} & \multicolumn{4}{c|}{Counts (complete / random test set)}\\ 
     & & \multicolumn{1}{c|}{TF-IDF} & \multicolumn{1}{c|}{Word2vec} & \multicolumn{1}{c|}{Transformer}  & \multicolumn{1}{c|}{RippleNet} \\ \hline
    
    \multirow{5}{*}{Q1} & (C1) & 600 / 81 & 575 / 68 & 575 / 88 & 470 / 83 \\ 
    & (C2) & 358 / 63 & 367 / 70 & 374 / 52 & 471 / 61 \\   
    & (C3) &  14 / 1  & 30  / 7  & 23  / 5  & 31/ 1 \\   
    & (C4) & 198 / 31 & 198 / 31 & 194 / 29 & 193/ 31 \\   
    & (C5) &   4 / 1  &  4  /  1 & 8  / 3   & 9 / 1 \\   \hline
    
    \multirow{5}{*}{Q2} & (C1) & 595 / 79 & 575 / 69 & 572 / 86 & 479 / 82 \\ 
    & (C2) &  362 / 65 & 367 / 69 & 378 / 54 & 462 / 62\\   
    & (C3) &   15 / 1  & 30  / 7  & 22  / 5  & 31 / 1\\   
    & (C4) &  199 / 31 & 198 / 31 & 193 / 29 & 194 / 31\\   
    & (C5) &   3  / 1  &  4  / 1  & 9   /  3 & 8 / 1\\   \hline
    
    \multirow{5}{*}{Q3} & (C1) & 600 / 78 & 574 / 68 & 574 / 86 & 477 / 81\\ 
    & (C2) & 360 / 66 & 370 / 70 & 375 / 54 & 467 / 63\\   
    & (C3) &  15 / 1  &  31 / 7  & 26  / 5  & 31 / 1\\   
    & (C4) & 196 / 31 & 195 / 31 & 193 / 30 & 191 / 31\\   
    & (C5) &   3 / 1  & 4   / 1  & 6   /  2 &  8 / 1\\   \hline
    
    \multirow{5}{*}{Q4} & (C1) & 611 / 82 & 590 / 76 & 538 / 86 & 495 / 87\\ 
    & (C2) &  342 / 54 & 356 / 61 & 402 / 48 & 448 / 50\\   
    & (C3) &   13 / 2  & 20  / 1  & 26  / 4  & 23 / 1\\   
    & (C4) &  206 / 38 & 198 / 39 & 204 / 37 & 200 / 39\\   
    & (C5) &   2  / 1  & 10  /  0 & 4  /  2  & 8 / 0\\   \hline
    
    \multirow{5}{*}{Q5} & (C1) & 624 / 89 & 592 / 77& 542 / 88 & 498 / 84\\ 
    & (C2) & 337 / 51 & 357 / 62 & 403 / 49 & 451 / 57\\   
    & (C3) &  10 / 2  &  22 / 3  & 26  / 5  & 22 / 1\\   
    & (C4) & 200 / 34 & 194 / 35 & 201 / 34 & 195 / 35\\   
    & (C5) &   3 / 1  & 9   /  0 & 2   /  1 &  8 / 0\\   \hline

    \end{tabular}%
    }
    \label{tab:counts-stance-bias}
\end{table}

Lastly, we compute the Pearson correlation coefficient between the recommender and the user average bias scores, to further investigate whether our initial observations are statistically significant. Tables \ref{tab:correlation-sentiment-bias} and \ref{tab:correlation-stance-bias} illustrate the results of recommender-user sentiment, and respectively, stance bias correlation. 

\begin{table}[!htb]    
    \centering
    \caption{Recommender-user sentiment bias correlation. Statistical significance is denoted with * ($p-value < 0.01$) and ** ($p-value < 0.05$ ).}
    \begin{tabular}{|l|r|r|r|r|}
    \hline
    \multirow{2}{*}{Test set} & \multicolumn{4}{c|}{Pearson Correlation (p-value)}\\ 
     & \multicolumn{1}{c|}{TF-IDF} & \multicolumn{1}{c|}{Word2vec} & \multicolumn{1}{c|}{Transformer}  & \multicolumn{1}{c|}{RippleNet} \\ \hline
    
    Complete & 0.325* & 0.347* & 0.216* & 0.045**  \\ \hline
    Random &  0.253* & 0.310* & 0.129 &  -0.033  \\   \hline
    
    \end{tabular}

    \label{tab:correlation-sentiment-bias}
\end{table}

\begin{table*}[!htb]    
    \centering
    \caption{Recommender-user stance bias correlation. Statistical significance is denoted with * ($p-value < 0.01$) and ** ($p-value < 0.05$ ).}
    \resizebox{0.6\textwidth}{!}{%
    \begin{tabular}{|l|r|r|r|r|}
    \hline
    \multirow{2}{*}{Question} & \multicolumn{4}{c|}{Pearson Correlation (p-value) (complete / random test set)}\\ 
     & \multicolumn{1}{c|}{TF-IDF} & \multicolumn{1}{c|}{Word2vec} & \multicolumn{1}{c|}{Transformer} & \multicolumn{1}{c|}{RippleNet} \\ \hline
    
    (Q1) & 0.248* / 0.078 & 0.196* / 0.029 & 0.187* / 0.162** & -0.022  / 0.111 \\ \hline
    (Q2) & 0.241* / 0.078 & 0.196* / 0.019  & 0.176* / 0.148** & -0.008  / 0.115  \\   \hline
    (Q3) & 0.252* / 0.072  & 0.199*  / 0.003  & 0.195* / 0.143 & -0.009  / 0.091  \\ \hline
    (Q4) & 0.280* / 0.135 & 0.249* / 0.002  & 0.143* / 0.230* & -0.019  / 0.148  \\ \hline
    (Q5) & 0.320* / 0.192** & 0.269* / 0.042  & 0.168* / 0.230* & -0.011  / 0.112 \\ \hline
    
    \end{tabular}%
    }
    \label{tab:correlation-stance-bias}
\end{table*}

In the case of \textit{sentiment bias}, the results indicate a statistically significant positive correlation between the recommender and the user bias for the TF-IDF and Word2vec models on both test sets. In the case of the Transformer recommender, we notice a statistically significant positive correlation only on the complete test set. This means that if a user prefers only articles with a negative sentiment towards the topic of, for example, refugees living in Germany, these recommenders will continue recommending news with a negative sentiment. In this case, the user's exposure to articles with positive sentiment will be constrained by the recommender, and the user will not have access to news that might not agree with his existing views. In contrast, in the case of RippleNet, we found a statistically significant correlation between the recommender and the user bias for the complete test set, for $p-value < 0.05$, and no statistical significance on the random test. The latter could potentially indicate that the recommender is less prone to amplifying the user's existing preference for a particular sentiment.

As far as \textit{stance bias} is concerned, text-based recommenders are also statistically significant positively correlated with the existing user bias on the complete test set, for all questions, as shown in Table \ref{tab:correlation-stance-bias}. However, this pattern does not hold true in the case of the random test set. In this case, we noticed that only the Transformer recommender shows a positive correlation with the user bias, which is statistically significant for four out of five questions, whereas the Word2vec model has no statistically significant bias correlation with the user history. In comparison, our analysis shows that the stance bias of the TF-IDF model is positively correlated with the user's bias only on question (Q5) for $p-value < 0.05$. Lastly, as in the case of sentiment bias, RippleNet shows no statistically significant correlation with the existing stance bias of users on either of the two tests sets, again demonstrating that the knowledge-based recommender is less prone to amplifying user bias.

\section{Discussion}
\label{sec:discussion}
In this paper, we have introduced GeNeG, a German news corpus on the topic of refugees and migration, including polarity annotations with respect to sentiment and stance. We have demonstrated that those annotations can be utilized to quantify the sentiment and stance bias of different recommender algorithms.

Our experiments show that purely text-based recommender systems expose amplification of user attitudes with respect to sentiment and stance, both constituting cues for the creation of filter bubbles in the process of algorithmic news curation, and the decrease of news diversity in terms of sentiments and viewpoints. Similar results have been found by Gao et al. \cite{GaoXKF18}, who conducted experiments to understand the effects of stance and credibility labels on online news selection and consumption. The results of their study show that these labels intensify the tendency of the people to look for opinions which are in-line with their own beliefs, and may lead the people being more vulnerable to polarized opinions and fake news.

In contrast, the knowledge-aware recommender appears to be less prone to both types of biases, but also less powerful in terms of accuracy on click-through rate prediction when computing recommendations solely based on entities. On the one hand, text-based recommenders are predisposed to identifying sentiment and stance cues in the text, as their recommendations are based only on textual information. On the other hand, using knowledge bases as sources of side information and taking into account solely named entities identified in the text (regardless of the textual context in which they are mentioned) diminishes such underlying bias, while capturing deeper knowledge-level connections between articles. Therefore, we argue that future research should find a balance between two obviously conflicting goals: performance (as understood by prediction accuracy) and bias (in terms of the sentiments and stances expressed in the recommended news articles). The results of this paper indicate that knowledge-based approaches are a good candidate when it comes to develop less biased recommenders. However, future work should investigate also the behavior of collaborative-filtering approaches in terms of sentiment and stance bias. Moreover, the analysis could be extended to other datasets in order to examine if the same patterns can be observed for a larger variety of topics.

As a future perspective, we would like to enable the recommender systems to take into account the sentiments, as well as the stance of an article. The news recommender system should be able to recommend topics which allow the user to consider differing opinions as compared to their own beliefs, hence reducing the effect of filter bubbles. Since in this paper we have shown how bias can be quantified, it would also be possible to introduce a combined score as a new optimization goal for algorithmic recommenders, which trades off accuracy and bias in a single score (just like, e.g., F1 score trades off recall and precision). Optimizing news recommender systems towards such a target function could be one possible approach towards developing less biased news recommender systems.

\begin{acks}
The work presented in this paper has been conducted in the ReNewRS project, which is funded by the Baden-Württemberg Stiftung in the Responsible Artificial Intelligence program.
\end{acks}

\bibliographystyle{ACM-Reference-Format}
\bibliography{references}

\appendix

\section{User Study and Data Collection}
\label{sec:user_study}

The experimental setup is extensively described in \cite{ludwig2022polarization}. The participants were recruited through the online-access panel of Respondi AG, and selected using a quote procedure to match German Internet users aged 18 to 74. The mean average age was 46.56, with a standard deviation of 15.68, 47.7\% of the participants were females, and 38.5\% of the respondents had an \emph{Abitur} (German university entrance certificate) or higher qualification. Only participants with desktop-based devices were included given the display of the experimental conditions. Moreover, some participants were removed from the final user dataset if these did not meet the age range of the quota sample, spent less than 120 seconds or more than 2.5 hours with the recommender system, or provided very consistent answering patterns across all constructs, resulting in a final sample of 1,417 participants out of 1,801 initial ones.

The study consisted in a two-factorial experiment with an incomplete design. The first factor consisted in using one of three types of text-based news recommender systems (i.e. TF-IDF, Word2vec, Transformer) or a random recommendation. The second factor concerned the inclusion of sentiment in the recommender. Only the TF-IDF recommender was enriched with sentiment scores, and thus, ratings gathered based on it are over-represented in the data. More specifically, in the test data, there are 786 unique users with TF-IDF generated ratings, 211 with Word2vec-generated ratings, 209 with Transformer-generated ratings, and 211 with randomly generated ratings.

Each participant was firstly asked to complete a questionnaire including questions regarding polarization, political interests, attitudes towards refugees, and demographics. Afterwards, the subjects were asked to select the news that they found the most interesting from a list of six news previews, consisting of their titles and first lines of text. The participants would then read the news, and the process would be repeated four times. At the end of the experiment, the participants would be asked again to complete a follow-up questionnaire, in order to examine how the recommended news influenced their political polarization measured by the initial questionnaire.

\end{document}